\documentstyle[twoside,fleqn,espcrc2]{article}
\input{psfig}
\newcommand{\AmS}{{\protect\the\textfont2
  A\kern-.1667em\lower.5ex\hbox{M}\kern-.125emS}}

\title{Duality in Josephson Junction Arrays}

\author{Ya.~M.~Blanter\address{D\'epartement de 
Physique Th\'eorique, Universit\'e de
Gen\`eve, CH-1211 Gen\`eve 4, Switzerland},
Rosario Fazio\address{Istituto di Fisica, Facolta di Ingegneria, 
Universit\`a di Catania, 95128 Catania, Italy},
and
Gerd Sch\"on\address{Institut f\"ur Theoretische Festk\"orperphysik,
Universit\"at Karlsruhe, 76128 Karlsruhe, Germany}
}
       
\begin{document}

\begin{abstract}
Various properties of mesoscopic 
two-dimensional Josephson junction arrays 
are reviewed. Particular attention is paid to 
structure of the topological excitations,
charges and vortices, which are shown to be dual to each other. 
This duality persists in the presence of external magnetic fields and 
offset charges, which influence vortices and charges in an equivalent
way. A double-layer junction array is also considered, 
where an even further reaching duality is discovered.  
\end{abstract}

\maketitle

\section{Introduction}

Since the original work by Kramers and Wannier~\cite{Kramers} on the  
two-dimensional (2D) Ising model, duality has been proven to be a 
powerful tool in field theory
and statistical mechanics~\cite{Savit}. The idea behind this
transformation is the mapping of the weak coupling region of the
system under consideration onto the strong coupling range (and vice
versa). The symmetries 
of the system under this transformation lead to important insight into the
structure of the model, especially in the intermediate region of coupling
constants which is usually elusive to standard treatments.
Dual transformations applied to the topological
excitations of the system are particularly advantageous since 
it is possible to recast the partition 
function solely in terms of these degrees of freedom. Jos\'e {\em et
al.} \cite{JKKN} applied these techniques to show that the 
vortices in 2D XY-model
can be mapped onto the charges of a two-dimensional Coulomb gas.
Kadanoff \cite{Kadanoff} showed that this mapping 
is not restricted to the XY-model, but various systems can be 
mapped onto coupled Coulomb gases. A review of these techniques 
applied to a  number of systems can be found in
Ref.~\cite{Nienhuis}.  
  
In the past years two-dimensional Josephson junction arrays (JJA) have
proven to be an excellent arena for the study of a variety  of phase 
transitions~\cite{MooijSchon}. They are fabricated from an array of 
superconducting islands connected by Josephson links. Each island is 
characterized by the modulus and the phase of the 
order parameter, $\Delta \exp (i\phi)$. Upon lowering the
temperature, each island of the array goes superconducting at the 
BCS critical temperature $T_{\rm c0}$. But, in spite of the fact that
each island is superconducting, the whole array remains in a resistive
state as long as the phases, driven both by thermal and quantum
fluctuations, have not acquired (sufficient) long range order. 
This global phase coherence sets in at a lower temperature. 
It is reasonable to assume that at these temperatures the magnitude
of the order parameter is fixed to its equilibrium value, 
and the array can be described only in terms of the phases $\phi_i$. 
The resulting Hamiltonian is
\begin{equation} \label{2DXY}
	H = - E_{\rm J} \sum_{\langle i j \rangle} 
		\cos(\phi_i - \phi_j) \; ,
\end{equation}
where $E_{\rm J} > 0$ is the Josephson coupling
energy, and the summation is over nearest neighbors.
The lattice model defined in Eq.(\ref{2DXY}) is equivalent to the 
2D XY-model -- a planar lattice of localized planar magnetic momenta
${\bf S_i}$ of unit length. It describes other physical systems 
as well, including planar ferromagnets, two-dimensional crystals~\cite{KT} 
or two-dimensional superconducting films \cite{HN}. 

The XY-model undergoes a Berezinskii-Kosterlitz-Thouless
(BKT) transition~\cite{Ber,KT,HH} driven by fluctuations in
the vortex density.
For the JJA this implies that the array is phase coherent (superconducting)
below the temperature $T_{\rm J}$, which is of the order of 
$E_{\rm J}$ (we use units where $\hbar = k_{\rm B} = 1$). 
This is demonstrated by a simple argument due to Kosterlitz and
Thouless. At zero temperature the lowest energy state of the system has 
the phases $\phi_i$ aligned, however, at finite temperatures  the
free energy $F = E - TS$ should be minimized. 
Consider an isolated vortex configuration. 
Its energy is readily calculated from Eq. (\ref{2DXY}).
Logarithmic divergences are cut off at 
short distances by the lattice spacing $a$ and at long distances 
by the system size $R$, hence $E = \pi E_{\rm J} \ln (R/a)$.
The entropy  of the vortex configuration is given by the logarithm of
the number of possible positions of the vortex, $S = 2\ln(R/a)$. Thus,
the free energy of the vortex becomes negative at the temperature 
$T_{\rm J} = \pi E_{\rm J}/2$. Below this temperature 
the phases are ordered, and vortices may appear only in bound
pairs. I.e. the XY-model is in a
ferromagnetic state, while the JJA is superconducting.
Above the transition temperature  $T_{\rm J}$ free
vortices may form. Their motion 
causes dissipation, and hence the state is resistive. 
This analysis is confirmed by the renormalization group calculations
\cite{Kosterlitz}. A thorough review concerning the statics and
dynamics of the BKT transition can be found in Ref. \cite{Minn}.

\section{Charge-vortex duality in 2D Josephson junction arrays}

If the superconducting islands are of sub-micron size, as can be realized
with modern nano-lithography, the electrostatic charging energy associated
with non-neutral configuration of the islands cannot be disregarded
any longer. It can be expressed as 
\begin{equation} \label{chclass}
	E_{\rm ch} = \frac{1}{2} \sum_{ij} V_i C_{ij} V_j.
\end{equation}
Here the summation is over all islands $i$ with local voltage
$V_i = \partial_t (\phi_i/2e)$. 
The capacitance matrix is determined in a good approximation by
the on-site and  the nearest neighbors elements, $C_{ii} = C_0 +
4C$, $C_{ij} = -C$ if $i$ and $j$ are nearest-neighbors, and $C_{ij} =
0$ otherwise. Here $C_0$ is the capacitance of the grains to the
ground while $C$ is the junction capacitance. 

In a quantum mechanical treatment, the charge $\hat Q_i$ on the island $i$
has to be viewed as conjugated to the phase $\phi_i$ (see e.g. 
Ref. \cite{SZ}), i.e.
\begin{equation}
	\hat Q_i = \frac{1}{i} \frac{\partial}{\partial 
                   (\phi_i/2e)}.
\end{equation}
Then the Hamiltonian of the array is \cite{Fazio}
\begin{equation}
		\hat H  = \frac{1}{2} \sum_{ij} 
			\hat Q_i C^{-1}_{ij} \hat Q_j -
			E_{\rm J} \sum_{\langle ij \rangle} 
			\cos(\phi_i - \phi_j).
\end{equation}
The interaction between the charges in the array is 
determined by the inverse capacitance matrix $C^{-1}_{ij}$.
Its range depends on the ratio $C/C_0$; if the 
capacitance to ground is much larger than the junction capacitance 
the charges feel an on-site interaction. In the opposite limit the Coulomb
interaction depends logarithmically on the distance. The capacitance
to the ground $C_0$ leads to a screening on the length scale 
 $\sim a \sqrt{C/C_0}$.

The two characteristic energy scales in the system are the Josephson  
coupling energy 
$E_{\rm J}$, which is associated to the  tunneling of Cooper pairs between 
neighboring islands, and the charging energy 
$E_{\rm C} = e^2 /(2\max\{ C_0, C\})$,
which is the energy cost to add an extra Cooper pair on a neutral
island. The charging energy  $E_{\rm C}$ tends to inhibit Josephson 
tunneling and is responsible for
quantum fluctuations of the phases $\phi_i$ on each island.
If $E_{\rm J} \gg E_{\rm C}$  the fluctuations of the phases are weak,
and the system acquires global superconducting coherence. 

In this section we derive some properties of quantum JJA from
duality arguments. Our starting point is the partition function,
\begin{eqnarray} \label{part1}
	Z = & & \prod_{i} \int_{0}^{2\pi} d\phi_i^{(0)} 
		\sum_{\{m_i = 0,\pm1,\dots\}} 
	\nonumber\\
 	& & \int_{\phi_{i}^{(0)}}^{\phi_{i}^{(0)} + 2\pi m_{i}} 
	D\phi_i (\tau) \exp (-S\{\phi\}).
\end{eqnarray}
Here the Euclidean 
path integration is carried out with the boundary conditions
\begin{equation} \label{bound1}
	\phi_{i} (0) = \phi_{i}^{(0)}; \ \ \ \phi_{i} (\beta) =
	\phi_{i}^{(0)} + 2\pi m_{i}, 
\end{equation}
with $\beta$ being the inverse temperature.
These non-trivial boundary conditions express the
fact that the charges of the grains are integer multiples of $2e$
\cite{Bound,SZ}. The Euclidean effective action $S\{\phi\}$ has the form
$$	S\{\phi\} =
	\int_{0}^{\beta} d\tau \left\{ \vphantom{\frac{}{{}_{}}}\right. 
	\frac{C_{0}}{8e^2}\sum_i
	(\dot{\phi}_{i})^2 
	+  \frac{C_1}{8e^2}\sum_{\langle ij \rangle} 
	(\dot{\phi}_{i} - \dot{\phi}_{j})^2 
$$
\vspace{-3mm}
\begin{eqnarray} \label{Sphi5}	
\;\;\;\;\;\;\;\;\;\;\;\;\;\;\;\;\;\;\;\;\;\;\;\;\;\;\;\;
	- E_{\rm J}\sum_{\langle ij \rangle} 
	\cos(\phi_{i} - \phi_{j}) \left.  \vphantom{\frac{}{{}_{}}} \right\}.
\end{eqnarray}

The model has been  studied extensively
to derive the phase diagram by means of mean
field~\cite{Abeles,Efetov,Simanek,Doniach,LozAk}, Monte
Carlo~\cite{Jacobs}, variational~\cite{Fishman} and RG~\cite{Jose1}
calculations. Non-perturbative features of this model and its connections
to planar Chern-Simons gauge theories were discussed in
Ref.~\cite{Diamantini}. Quantum fluctuations lower the temperature of the 
BKT phase transition separating resistive and 
superconductive phases~\cite{Jose1}
below   the transition temperature $T_{\rm J0}$ of the classical array 
(with $E_{\rm C} = 0$). In the case $C_0 = 0$ the shift of the
transition temperature is
\begin{equation}
	T_{\rm J} = T_{\rm J0} - \frac{e^2}{48\pi C}.
\end{equation}
Beyond a critical value of the charging energy
$E_{\rm C}$ the transition temperature vanishes, 
and the array remains insulating even at zero temperature.

The purpose of this Chapter is to show that there is a dual transformation
relating the classical limit, $E_{\rm J} \gg E_{\rm C}$, 
to the opposite quantum limit, $E_{\rm J} \ll E_{\rm C}$ 
\cite{Fisher,Mooij2,Fazio,FGS}. In the latter
case the quantum fluctuations of the phases are strong,
and vortices are ill-defined objects. However, in this regime 
the charges on the islands are well-defined variables. 
In the extreme limit $C_0 \ll C$, the interaction between the charges
is logarithmic, in the same way as that of the vortices in the classical
array. In this case the charges form a 2D Coulomb 
gas and undergo a BKT transition at temperature $T_{\rm C} \sim
E_{\rm C}$, where the state below the transition is insulating. If the
capacitance to the ground is larger, no finite-temperature
phase transition arises. However, we can still expect a crossover from a
low-temperature phase with exponentially low conductance
to a resistive phase at finite temperature.

In order to describe an intermediate situation we need a formulation
in terms of both charges and vortices.
Following Ref.~\cite{Fazio,FGS} we first introduce 
the island charges in a path integral
representation. In terms of phase trajectories $\phi_i(\tau)$ and
charges $q_i(\tau) = Q_i(\tau)/2e$ the partition function takes the form
\begin{eqnarray} \label{part11}
	Z  =  \prod_j \int dq_{j0} \int Dq_j(\tau) \prod_{i} \int_{0}^{2\pi}
	 d\phi_i^{(0)} \sum_{\{m_i\}} \nonumber\\
	  \int D\phi_i (\tau) \exp (-S\{\phi,q\}),
\end{eqnarray}
where the phases obey the boundary conditions (\ref{bound1}),
while the charge paths are periodic, 
$q_j(0) = q_j(\beta) = q_{j0}$. 
The effective action in terms of phases $\phi_i(\tau)$ and charges
$q_i(\tau)$ is
$$
	S\{\phi,q\} = \int_{0}^{\beta} d\tau 
	\left\{  \vphantom{\frac{}{{}_{}}}\right.
	2e^2 \sum_{i,j} q_i  C^{-1}_{ij} q_j   
	\;\;\;\;\;\;\;\;\;\; \;\;\;\;\;\;\;\;\;\;
$$
\vspace{-0.4cm}
\begin{eqnarray} \label{Sqphi5}
\;\; \;\;\;\;\;\;\;\;\;\; + i \sum_i q_i  \dot\phi_i  
	- E_{\rm J}\sum_{\langle ij \rangle} 
	\cos (\phi_{i}  - \phi_{j} )
				\left.	 \vphantom{\frac{}{{}_{}}} \right\}. 
\end{eqnarray}
The summation over winding numbers $\{m_i\}$ fixes the
charges $q_i$ to be integer-valued \cite{Bound}. 

Following the steps discussed in the Appendix, the
partition function can be expressed as a sum over charge 
and vortex configurations\\

\begin{equation} \label{part12}
	Z = \sum_{\{q_{i\tau}\}}\sum_{\{v_{i\tau}\}} 
	\exp\left(- S \{q,v\}\right) \; .
\end{equation}
The effective action for the integer charges $q_i(\tau)$ and vorticities
$v_i(\tau)$ is 
$$
	S\left\{ q, v \right\} =  
	\int_0^{\beta} d\tau \sum_{ij} \Big\{ 2e^2 q_i 
	C^{-1}_{ij} q_j 	+ \pi E_{\rm J} v_i  G_{ij} v_j    
$$
\vspace{-0.4cm}
\begin{equation} \label{Sqv11}
  \;\;\;\;\;\;\;\;\;\; \;\;\;\; 
+ \, iq_i  \Theta_{ij} \dot{v}_j  
	+  \frac{1}{4\pi E_{\rm J}}
	\dot{q}_i  G_{ij} \dot{q}_j  \Big\}.
\end{equation}
It describes two coupled Coulomb gases. The
charges interact via the inverse capacitance matrix. The interaction
among the vortices is described by the kernel
$G_{ij}$, which is obtained as the Fourier transform of $k^{-2}$. 
At large distances  $r_{ij} \gg a$ between the
sites $i$ and $j$ it depends logarithmically on the distance 
\begin{equation}
	G_{ij} = \frac{1}{2}\ln\left(\frac{a}{r_{ij}}\right) \; .
\end{equation} 
The charges and vortices are coupled in the dynamical theory 
by the third term. Here  
\begin{equation}
\Theta_{ij} = \arctan\left(\frac{y_i - y_j}{x_i - x_j}\right)
\end{equation}
describes the phase configuration at the site $i$ if a vortex is placed 
at the site $j$. The coupling has a simple physical interpretation: a
change of vorticity at site $j$ produces a voltage at site $i$ which
is felt by the charge at this location. The last term 
$\dot{q} G \dot{q}$ represents a
spin-wave contribution to the charge correlation function. 

The effective action (\ref{Sqv11}) shows a high degree of symmetry
between vortex and charge degrees of freedom. In particular, in the
limit $C_0 \ll C$ the inverse capacitance matrix depends on distance
in the same way as the vortex interaction,
\begin{equation}
	e^2 C^{-1}_{ij} = \frac{E_{\rm C}}{\pi} G_{ij} \; ,
\end{equation}
and charges and vortices are (nearly) dual. The duality is
broken by the last term $\dot{q} G \dot{q}$. This term is ``irrelevant''
for the phase transitions, i.e.\ it merely shifts the transition
point.
But it has important consequences for the dynamical behavior.

The action (\ref{Sqv11}) is written in terms of integer charges and
vorticities. However, depending on the coupling constants, only one type 
of excitations may be well defined. In the quasiclassical limit, 
$E_{\rm J} \gg E_{\rm C}$, vortices are the relevant excitations, 
while charges are strongly fluctuating and  can be treated as continuous
variables. By integrating out the charges, one obtains
the effective action for vortices $v_i(\tau)$
$$
	S\{v\} = \int_0^{\beta} d\tau \Big\{ \frac{1}{8e^2} \sum_{ijkl}
	\dot{v}_{i} \Theta_{ik} C_{kl} \Theta_{lj} \dot{v}_{j}
\;\;\;\;\;\;\;\;\;\;\;\;\;\;\;\;\;	
$$	
\vspace{-5mm}
\begin{equation} \label{Sv1}
\;\;\;\;\;\;\;\;\;\;\;\;\;\;\;\;\;\;\;\;\;\;\;\;\;\;\;\;\;\;\;\;\;\;	 
	+ \pi E_{\rm J} \sum_{ij} v_{i} G_{ij} v_{j} \Big\}.
\end{equation}
In the limit of low self-capacitances, $C_0 \ll C$, the kernel in
the first term becomes $\Theta_{ik} G^{-1}_{kl} \Theta_{lj}$,
and the effective action for vortices
reduces to \cite{NH}
$$
S\{v\} = \int_0^{\beta} d\tau  \sum_{ij} \Big\{ \frac{\pi}{8E_{\rm C}} 
	\dot{v}_{i} G_{ij} \dot{v}_{j}
\;\;\;\;\;\;\;\;\;\;\;\;\;\;\;\;\;	\;\;\;\;\;\;\;\;\;\;\;	
$$	
\vspace{-5mm}
\begin{equation} \label{Sv2}
\;\;\;\;\;\;\;\;\;\;\;\;\;\;\;\;\;\;\;\;\;\;\;\;\;\;\;\;\;\;\;\;\;\;	 
	+ \pi E_{\rm J} \sum_{ij} v_{i} G_{ij} v_{j} \Big\} \; .
\end{equation}
The summation in the partition function
is constrained by the neutrality condition $\sum_i v_i = 0$. 

The effective action (\ref{Sv2}) describes a {\it quantum
Coulomb gas} of vortices. This becomes clearer if we 
change from a description in terms of the vorticity at site $i$ 
 to a continuous description where we
label the vortices by their center coordinate ${\bf r}(\tau)$ 
and sign of vorticity $v_n = \pm 1$ \cite{Foot1}. Both are
related by 
\begin{eqnarray} 
	v_{i}(\tau) \to \sum_{n} v_n \delta({\bf r}_i - {\bf r}_n(\tau))\; .
\end{eqnarray}
 In these variable the partition function becomes,
\begin{equation} \label{part31}
	Z = \sum_{N=0}^{\infty} \int D{\bf r}_1 (\tau) \dots 
	{\bf r}_{2N}(\tau) \exp(-S\{ {\bf r} \}) \;. 
\end{equation}
(We have explicitly used the charge neutrality conditions: the
integration is carried over $N$ vortices with $v=1$ and $N$
anti-vortices with $v=-1$). The effective action expressed in terms of
the vortex coordinates is
$$
	S\{ {\bf r} \} = \int_0^{\infty} d\tau \sum_{m,n=1}^{2N} \Big[
	\frac{1}{2} \dot{r}^{\alpha}_m M_{\alpha\beta}
	({\bf r}_m - {\bf r}_n) \dot{r}^{\beta}_n \;\;\;\;\;\;\;\;\;\;
$$
\begin{equation} \label{Sr}
\;\;\;\;\;\;\;\;\;\;\;\;\;\;\;\;\;\;\;\;\;\;\;\;\;\;\;\;\;\;\;\;\;\;
	+ \pi E_{\rm J} G({\bf r}_m -{\bf r}_n) \Big].
\end{equation}
The second term in (\ref{Sr}) 
is the interaction energy of the vortices.
In the first term we introduced the vortex mass tensor \cite{vmass},
\begin{equation} \label{mass11}
	M_{\alpha\beta} (\mbox{\bf r}) 
	= -\frac{\pi}{4E_{\rm C}} \nabla_{\alpha} \nabla_{\beta}
	G(\mbox{\bf r}). 
\end{equation}
For $r \gg a$ it decreases as $r^{-2}$, and consequently may
be approximated by a local function
\begin{equation} 
	M_{\alpha\beta} (\mbox{\bf r}) = M_{\rm v} \delta_{\alpha\beta}
\delta(\mbox{\bf r}), \ \ \ 
	M_{\rm v} = \frac{\pi^2}{4a^2 E_{\rm C}}\; . 
\end{equation}
It defines the vortex mass $M_{\rm v}$~\cite{vmass,ES}.
In this case the first term represents a
kinetic energy of vortices, which can be rewritten as
\begin{equation} \label{s311}
	S_{\rm kin}^{\rm (v)} =  \int_0^{\beta} d\tau
	\sum_{m=1}^{2N} \frac{1}{2} M_{\rm v} \dot{r}_m^2 (\tau) \; .
\end{equation}

A similar consideration for $E_{\rm C} \gg E_{\rm J}$ 
lead us to the action of a 2D
Coulomb gas of charges with charge mass 
$M_{\rm q} = (a^2 E_{\rm J})^{-1}$ (see also Ref. \cite{Hermon}). 
Thus, in two dimensional arrays of Josephson junctions a charge-vortex 
duality exists. In the limit $E_{\rm J} \gg E_{\rm C}$ the 
vortices are well-defined. They form a Coulomb gas, and can be considered as
particles with masses. In the opposite limit 
$E_{\rm C} \ll E_{\rm J}$ the charges
are the relevant excitations. The charges have the same  properties as
the vortices in the corresponding limiting cases.

\section{Double-layered arrays}

Another system of increasing experimental interest is composed of 
two parallel 2D Josephson junction arrays with purely 
capacitive coupling between them (no Josephson
coupling~\cite{Foot2}). We will show that in this case an even further
reaching duality between charges and vortices arises. We restrict 
ourselves to the most interesting situation,  when one array is in the 
quasi-classical (vortex) regime while the other is in the quantum 
(charge) regime. 
Then the vortices in one layer and the charges in the other
 are simultaneously well-defined {\em dynamical} variables. 
(In contrast in a single array vortices or charges acquire a kinetic energy
only after the other variable is integrated out.)
Another important
feature of the present system is that the strength of interaction
between charges and vortices is controlled by the interlayer
coupling $C_{\rm x}$ and consequently may be tuned independently.
We show that the physical realization of this interaction is rather
different from that in a single array. The theoretical description of this
system has been developed in Refs.~\cite{BS,Jose3}. 

The partition function of the system in
terms of the phases $\phi_{i\mu}$ (the indices $i$ label the islands
in each array and   $\mu = 1,2$ refers to the layer) is
$$
	Z = \prod_{i} \int_{0}^{2\pi} d\phi_{i1}^{(0)} d\phi_{i2}^{(0)}
	\sum_{\{m_{i1}\},\{m_{i2}\}} 
\;\;\;\;\;\;\;\;\;\;\;\;\;\;\;\;\;\;\;\;\;\;\;\;
$$
\begin{equation} \label{partphi}
\;\;\;\;\;\;\;\;\;\;
	\int D \phi_{i1} (\tau) D\phi_{i2}(\tau)
	\exp(-S\{\phi_1,\phi_2\})\; ,\;
\end{equation}
with boundary conditions
$\phi_{i\mu} (0) = \phi_{i\mu}^{(0)}$ and $\phi_{i\mu} (\beta) =
\phi_{i\mu}^{(0)} + 2\pi m_{i\mu}$. The Euclidean effective action
$S\{\phi_1,\phi_2\}$ has the form
$$
	 S\{\phi_1,\phi_2\} =
	\int_{0}^{\beta} d\tau \left\{  \vphantom{\frac{}{{}_{}}} \right.
	 \sum_{\mu=1,2}
	 \left[ \vphantom{\frac{}{{}_{}}} \right.
	\frac{C_{0\mu}}{8e^2}\sum_i
	(\dot{\phi}_{i\mu})^2  
\;\;\;\;\;\;\;\;\;\;\;\;\;\;\;\;\;\;\;\;\;\;\;\;
$$
$$ 
	+ \frac{C_{\mu}}{8e^2}
	\sum_{\langle ij \rangle} 
	(\dot{\phi}_{i\mu} - \dot{\phi}_{j\mu})^2 
	- E_{\rm J\mu}\sum_{\langle ij \rangle} 
		 \cos (\phi_{i\mu} - \phi_{j\mu})
	\left. \vphantom{\frac{}{{}_{}}} \right]
$$
\begin{equation} \label{Sphi}	
\;\;\;\;\;\;\;\;\;\;\;\;\;\;\;\;\;\;
	 +  \frac{C_{\rm x}}{8e^2}
	\sum_i (\dot{\phi}_{i1} - \dot{\phi}_{i2})^2 
		\left.  \vphantom{\frac{}{{}_{}}} \right\}.
\end{equation}
Here $C_{0\mu}$ are the capacitances of the islands in the array $\mu$
relative to the ground, $C_{\mu}$ are the capacitances of the
junctions in the
array $\mu$, and $C_{\rm x}$ are the interlayer capacitances
between adjacent islands, while $E_{\rm J\mu}$ are
the Josephson coupling constants in the layers. 

We concentrate on the situation
 in which the array 1 is in the quantum (charge)
regime while the array 2 is in the quasi-classical (vortex)
regime, i.e.
$$
	E_{\rm J1} \ll e^2/2\tilde C_1,\ \ \ E_{\rm J2} \gg e^2/2\tilde C_2 \;,
$$
with $\tilde C_{\mu} = \max \{C_{0\mu}, C_{\mu}, C_{\rm x} \}$. 
Various types of problems van be studied in these systems. 
If we are interested only in behavior of the vortex array (e.g. in an
experimental realization where one measures the quantities in the vortex
array only) we can integrate out all degrees of
freedom related to the charge array and study an effective action for
the vortex array. An analysis analogous to that presented for a single
array
shows that this action is essentially that of the Coulomb gas~\cite{BS}.
Consequently the vortex array undergoes a BKT transition, and in the
quasiclassical regime its temperature is lowered due to
electrostatic coupling to the charge array. 

A similar problem can be solved for the charge array. Then 
the charge-BKT transition changes into a crossover, since the
capacitive coupling to the vortex array leads to screening similar as
capacitances  to the ground. 

Our purpose here is to  consider both arrays simultaneously
and to investigate the charge-vortex duality in this system.
Similar as in a single array, we move from a description in terms of
phases to one in terms of charges and vortices, and use the duality of
the resulting action to investigate the transition. We will show that
charges and vortices in this system can 
be considered as two-dimensional dynamical particles with masses. The
charge-charge 
and vortex-vortex interaction are essentially those of 2D Coulomb
particles, while the charge-vortex interaction is highly interesting.

Before we proceed with the calculation it is necessary 
to stress the following. In
the regime of interest the interlayer capacitances $C_{\rm x}$ not
only couple the layers, but also renormalize the capacitances $C_{01}$ and
$C_{02}$ of the islands to the ground. The physical reason for this is
that due to the strong fluctuations of charges in layer 2 and
vorticities in layer 1 these variables are effectively continuous, and
hence a coupling to the other array plays the same role as a
coupling to the ground.
Hence the interaction between the charges in each layer has a
finite range for any non-zero $C_{\rm x}$ due to the screening, and the BKT
transition is replaced by a crossover. On the other hand,  in the
limit $C_{01} \ll C_{\rm x} \ll C_1$ the screening length $\xi_1 \sim
a(C_1/C_{\rm x})^{1/2}$ can be very large. Below we assume that
these inequalities are satisfied and the range of interaction $\xi_1$ is
large enough to make it meaningful to
speak about the charge-unbinding transition. This just means that the
crossover is (exponentially) sharp. For not so weak coupling $C_{\rm x}$
this description becomes meaningless, since the crossover is strongly
smeared, and the insulating phase is absent. 

To proceed it is convenient to introduce the large capacitance
matrix 
\begin{eqnarray} \label{matrix1}
	\hat C  = \left( \begin{array}{cc}
			\hat C_1 & -\hat C_{\rm x} \\
			-\hat C_{\rm x} & \hat C_2 \\
			\end{array} \right).
\end{eqnarray}
Here $\hat C_{\mu}$ is the capacitance matrix in the array $\mu$ while
$\big(\hat C_{\rm x}\big)_{ij} = C_{\rm x} \delta_{ij}$. The inverse matrix,
 describing the interaction of charges, in the Fourier
representation reads as
$$
\hat{C}^{-1} (\mbox{\boldmath k}) = 
	\left\{\prod_{\mu=1,2}\left(C_{0\mu} + C_\mu \mbox{\boldmath k}^2 
		+ C_{\rm x}\right)
%\left(C_2 \mbox{\boldmath k}^2 + C_{\rm x} + C_{02}\right)
	 - C_{\rm x}^2 \right\}^{-1}
$$
\begin{equation} \label{matrix2}
		\times \left(
	\begin{array}{cc}
		C_{02} + C_2 \mbox{\boldmath k}^2 + C_{\rm x} & C_{\rm x} \\
		C_{\rm x} & C_{01} + C_1 \mbox{\boldmath k}^2 + C_{\rm x}
	\end{array} \right) .  
\end{equation}
The effective action (\ref{Sphi}) can 
be rewritten in terms of integer charges $q_{i\mu}(\tau)$
and phases $\phi_{i\mu}(\tau)$   of each island 
$$
	S\{q,\phi\} = \int_{0}^{\beta} d\tau \left\{ 
		\vphantom{\frac{}{{}_{}}} \right. 2e^2 
	\sum_{ij,\mu,\nu} q_{i\mu}  \big(\hat C^{-1}\big)^{\mu\nu}_{ij}
	q_{j\nu}  \;\;\;\;\;\;\;\;\;\;
$$
\begin{equation} \label{Sqphi}
 	 + i\sum_{i\mu} 
	\Big[ q_{i\mu}  \dot{\phi}_{i\mu}  
	- E_{\rm J\mu}\sum_{\langle ij \rangle} 
	\cos (\phi_{i\mu} - \phi_{j\mu}) \Big]
	\left. \vphantom{\frac{}{{}_{}}} \right\}.
\end{equation}
It is possible to introduce vortex degrees of freedom in the same
way as for one array. Since
this procedure deals only with the phase variables and does not affect
the charge interaction (the first term in Eq. (\ref{Sqphi})),
the generalization for double-layered arrays is trivial. We
obtain 
\begin{equation} \label{z2}
	Z = \sum_{\{q_{i1}(\tau),q_{i2}(\tau)\}} 
	\sum_{\{v_{i1}(\tau),v_{i2}(\tau)\}} \exp (-S\{q,v\}), 
\end{equation}
where the effective action for integer charges $q_{i\mu}$ and vorticities
$v_{i\mu}$ is
$$
	 S\{q,v\} = \int_{0}^{\beta} d\tau  
	\left\{ \vphantom{\frac{}{{}_{}}} \right.
	2e^2 
	\sum_{ij,\mu,\nu} q_{i\mu}  \big(\hat C^{-1}\big)^{\mu\nu}_{ij}
	q_{j\nu}   \;\;\;\;\;\;\;\;\;\;\;\;
$$
$$ 	+ \sum_{\mu} \left[ \vphantom{\frac{}{{}_{}}} \right.
	\frac{1}{4\pi E_{\rm J\mu} 
	F(\epsilon_\mu E_{\rm J\mu})}  \sum_{ij}
	\dot{q}_{i\mu}   G_{ij} \dot{q}_{j\mu}   
$$
$$  + \, \pi E_{\rm J\mu} F(\epsilon_\mu E_{\rm J\mu}) 
		\sum_{ij} v_{i\mu} G_{ij} v_{j\mu} 
$$
\vspace{-0.4cm}
\begin{eqnarray} \label{Sqv1}
	 \;\;\;\;\;\;\;\;\;\;\;\; \;\;\;\;\;\;\;\;\;\;\;\; \;\;\;\;\;\;\;\;\;\;\;\;
	+ \, i\sum_{ij}  \dot{q}_{i\mu}  
	\Theta_{ij} v_{j\mu}   
	\left. \vphantom{\frac{}{{}_{}}} \right]
	\left. \vphantom{\frac{}{{}_{}}} \right\} .
\end{eqnarray}
Here we wrote terms which arise from the discretization of the
 time with lattice
spacing (in array $\mu$) of order $\epsilon_{\mu} \sim
(8E_{\rm J\mu}E_{\rm C\mu})^{-1/2}$ and $E_{\rm C\mu} \equiv
 e^2/2C_{\mu}$. (See the Appendix for the function $F$ 
renormalizing the
Josephson coupling). In the limit considered only the Josephson
 coupling of array 1 is noticeably renormalized, 
\begin{equation} \label{EJren}
	\tilde E_{\rm J1} \sim \left(8E_{\rm J1}E_{\rm C1}\right)^{1/2}
	\left[\ln(E_{\rm C1}/E_{\rm J1})\right]^{-1}.
\end{equation}
We suppress the tilde from now on. 
Later  we will  assume that the linear size of each
array is much less than the range of interaction $\xi_{\mu} =
a(C_{\mu}/C_{\rm x})^{1/2}$. This means, in particular, that we assume
$C_{\rm x} \ll C_2$.   

The action (\ref{Sqv1}) depends on the charges and vorticities in both
layers. However, in our situation, when the layers 1 and 2 are in the
charge and vortex regimes, respectively, the vortices in the layer 1
and the charges in the layer 2 are strongly fluctuating
degrees of freedom and
may be integrated out. To do this we suppose the latter
variables to be continuous and neglect the
spin-wave charge coupling term,
$\dot{q}_2G\dot{q}_2$,  in layer 2. 
Then after performing the Gaussian integration we obtain
the effective action for charges $q_{i1}(\tau)$ in layer 1 and
vorticities $v_{i2}(\tau)$ in layer 2 (to be referred below as $q_i$ and
$v_i$)
$$       S\{q,v\} = \int_{0}^{\beta} d\tau
\left\{ \vphantom{\frac{}{{}_{}}} \right. 2e^2
        \sum_{ij} q_i \big( \hat C_{ij}^{-1} \big)^{11}_{ij} q_j
 \;\;\;\;\;\;\;\;\;\;\;\; \;\;\;\;\;\;\;\;\;\;\;\;
$$
$$ + \frac{1}{4\pi E_{\rm J1}} \sum_{ij} \dot{q}_i G_{ij}
        \dot{q}_j
        + \pi E_{\rm J2} \sum_{ij} v_i G_{ij} v_j
$$
$$
        + \frac{1}{8Ee^2}
        \sum_{ijkl} \dot{v}_i \Theta_{ik} \Big[ \big( \hat C^{-1}
        \big)^{22} \Big]^{-1} _{kl} \Theta_{lj}
        \dot{v}_j
$$
\begin{eqnarray} \label{Sqv2a}
        + i \sum_{ijkl} \dot{v}_i
        \Theta_{ik} \Big[ \big( \hat C^{-1}
        \big)^{22}\Big]^{-1}_{kl} 
	\Big[ \big( \hat C^{-1} \big)^{12} \Big]^{-1}_{lj} q_ j
        \left. \vphantom{\frac{}{{}_{}}}\right\}.
\end{eqnarray}
In the limit where the charge interaction is long-range
we can use the appropriate limits of
the large capacitance matrix (\ref{matrix2}). In this case the action
reduces to
$$	 S\{q,v\} = \int_{0}^{\beta} d\tau 
\left\{ \vphantom{\frac{}{{}_{}}} \right. \frac{2E_{\rm C1}}{\pi}
	\sum_{ij} q_i G_{ij} q_j\;\;\;\;\;\;\;\;\;\;\;\;\;\;\;\;\;\;\;\;
$$
$$ + \frac{1}{4\pi E_{\rm J1}} \sum_{ij} \dot{q}_i G_{ij} 
	\dot{q}_j
	+ \pi E_{\rm J2} \sum_{ij} v_i G_{ij} v_j 
$$
$$
	+ \frac{\pi}{8E_{\rm C2}}
	\sum_{ij} \dot{v}_i \Big[ G_{ij} 
		- \frac{C_{\rm x}^2}{4\pi^2
	C_1C_2} \sum_{kl} \Theta_{ik} G_{kl} \Theta_{lj} \Big]
	\dot{v}_j
$$
\begin{eqnarray} \label{Sqv2}
\;\;\;\;\;\;\;\;\;\;\;\; \;\;\;\;\;\;\;\;\;\;
	+ \frac{iC_{\rm x}}{2\pi C_1} \sum_{ijk} \dot{v}_i
	\Theta_{ik} G_{kj} q_j
	\left. \vphantom{\frac{}{{}_{}}}	\right\}. 
\end{eqnarray}
This form displays the duality between charges and vortices in the
appropriate limit.

The action (\ref{Sqv2}) is the central result of this section. It
looks rather similar to the effective charge-vortex action in one
Josephson junction array. The most important difference is that
in one layer either charges or vortices are well-defined degrees
of freedom, while the action (\ref{Sqv2}) of the double-layer array
describes the system of two {\em well-defined} dynamic
variables on each site -- the charges in layer 1 and the vortices in 
layer 2. The action shows a duality 
between charges and vortices (the second term in
the square brackets is small for $C_{\rm x} \ll C_1,C_2$). Both kinetic
terms for charges and vortices violate the duality due to the
numerical coefficients. However, close enough
to the transitions these terms produce only small renormalization of
the transition temperature, and are irrelevant. 
Another interesting feature of this action is that the last term,
describing the 
interaction between charges and vortices, is also small, while in a
single-layer array the interaction is always of the same order of
magnitude as the other terms. 
 
It is obvious that for long-range interaction of the charges in 
layer 1 these also exhibit a BKT
transition, and under the conditions where the action (\ref{Sqv2}) was
obtained the transition temperature does not feel the presence of
layer 2. Hence
$$ 
	T_{\rm C} = T_{\rm C0}  - \frac{E_{\rm J1}}{24\pi} \; .
$$ 

To understand the physics described by the action (\ref{Sqv2}) it is
instructive to map this model onto the 2D Coulomb gas. For this purpose
we move again from the space-time lattice to the continuous medium and
introduce the coordinates of the vortex centers and charges
\begin{eqnarray} \label{decompose}
	& & q_i(\tau) \to \sum_{m} q_m \, \delta(\mbox{\bf r} 
	- \mbox{\bf r}_m(\tau)) \nonumber \\
	& & v_i(\tau) \to \sum_{n} v_n \, \delta(\mbox{\bf r} 
	- \mbox{\bf R}_n(\tau)) \; . 
\end{eqnarray}
Here $q_m = \pm 1$ and $v_n = \pm 1$ represent charges and
vortices, respectively, and $\mbox{\bf r}_m(\tau)$ and $\mbox{\bf
R}_n(\tau)$  the  coordinates of their
centers. In terms of these variables the partition function reads 
$$
Z = \sum_{M=0}^{\infty} \sum_{N=0}^{\infty} 
	\int D\mbox{\bf r}_1(\tau)
	\dots D\mbox{\bf r}_{2M}(\tau)\;\;\;\;\;\;\;\;\;\;\;\;\;\;\;\;\;\;\;\;
$$
\begin{equation} \label{part3}
\;\;\;\;\;\;\;\;\;\;	 D\mbox{\bf R}_1(\tau) \dots 
	D\mbox{\bf R}_{2N}(\tau) \exp(-S\{\mbox{\bf r},\mbox{\bf R} \}), 
\end{equation}
with an effective action $S\{\mbox{\bf
r},\mbox{\bf R} \}$ describing a neutral system of
 $2M$ positive and  negative charges ($q=\pm 1$)  
and of $2N$ positive and  negative vortices  ($v=\pm 1$).  

The first and third terms of the action (\ref{Sqv2}) can be easily
transformed by means of decomposition (\ref{decompose}). The first one
produces the potential energy of charge interaction, \\
\\
$
	S_{\rm int}^{\rm (q)} = 
$
\begin{equation} \label{s1}
	\frac{2E_{\rm C1}}{\pi} \int_0^{\beta} d\tau
	\sum_{m,n=1}^{2M} q_m q_n G (\mbox{\bf r}_m(\tau)
	 - \mbox{\bf r}_n(\tau)). 
\end{equation}
In principle, the summation includes the terms with $m = n$; these,
however, give rise only to the chemical
potential for charges. The third term in Eq. (\ref{Sqv2}) yields the
interaction of vortices,\\
\\
$
	S_{\rm int}^{\rm (v)} = 
$
\begin{eqnarray} \label{s2}
	\pi E_{\rm J2} \int_0^{\beta} d\tau 
	\sum_{m,n=1}^{2N} v_m v_n 
	G (\mbox{\bf R}_m(\tau) - \mbox{\bf R}_n(\tau)).
\end{eqnarray}
Here again the term with $m = n$ gives rise to the chemical potential
for vortices. The terms (\ref{s1}) and (\ref{s2}) are essentially the
action for (classical) Coulomb gases of charges and vortices,
respectively \cite{Kosterlitz}. 

If we neglect the small correction proportional to the $C_{\rm x}^2/C_1C_2$
in the fourth term in Eq.(\ref{Sqv2}) then the second and fourth terms
can be transformed to the kinetic energy of charges and vortices
respectively \cite{Fazio}. The second term gives
\begin{equation} \label{s31}
	S_{\rm kin}^{\rm (q)} = \frac{1}{2} \int_0^{\beta} d\tau
	\sum_{m,n=1}^{2M} q_m q_n \dot{r}_m^{\gamma} M_{\gamma\delta}
	(\mbox{\bf r}_m - \mbox{\bf r}_n) \dot{r}_n^{\delta},
\end{equation}
where the charge  mass tensor can be approximated as
\begin{eqnarray} \label{mass}
	M_{\gamma\delta} (\mbox{\bf r}) & = & - \frac{1}{\pi E_{\rm J1}} 
	\nabla_{\gamma} \nabla_{\delta}
	G(\mbox{\bf r}) \approx M_{\rm q} \delta_{\gamma\delta}
	\delta(\mbox{\bf r}) \nonumber \\ 
	M_{\rm q} & = & \frac{1}{a^2E_{\rm J1}}.
\end{eqnarray}
Then the kinetic term for charges takes a simple form
\begin{equation} \label{s3}
	S_{\rm kin}^{\rm (q)} = \int_0^{\beta} d\tau
	\sum_{m=1}^{2M}  \frac{1}{2} M_{\rm q} \dot{r}_m^2 (\tau)\; .
\end{equation}
Similarly, the fourth term in Eq.(\ref{Sqv2}) produces the
kinetic term for vortices
\begin{eqnarray} \label{s4}
	S_{\rm kin}^{\rm (v)} = & & 
	\int_0^{\beta} d\tau
	\sum_{m=1}^{2N} \frac{1}{2} M_{\rm v} \dot{R}_m^2 (\tau)
	\nonumber \\
	 \mbox{with} \; \;	& & 
	M_{\rm v} = \frac{\pi^2}{4 a^2 E_{\rm C2}}\; .
\end{eqnarray}
 
Finally, the last term in Eq.(\ref{Sqv2})) is responsible for the
interaction between charges and vortices. The corresponding term in  
the action is
$$
	S_{\rm qv} = \frac{iC_{\rm x}}{2\pi C_1 a^2} 
		\int_{0}^{\beta} d\tau \sum_{mn}v_m q_n
 \; \;\;\;\;\;\;\;\;\;\;\;\;\;\;\; \;\;\;\;\;\;\;\;\;\;\;\;
$$
\begin{equation} \label{s51}
	 \;\;\;\;\;\;
	  \int d\mbox{\bf r}' \nabla_{\mbox{\bf R}_m} 
		\Theta  (\mbox{\bf R}_m - \mbox{\bf r}') G
		(\mbox{\bf r}' - \mbox{\bf r}_n) 
		\dot{\mbox{\bf R}}_m(\tau)\; .\;
\end{equation}
After the integration over $\mbox{\bf r}'$ it reduces to
\begin{equation} \label{s5}
	S_{\rm qv} = - \int_{0}^{\beta} d\tau \sum_{m} i v_m 
	\dot{\mbox{\bf R}}_m(\tau) \mbox{\bf A} (\mbox{\bf R}_m)
\end{equation}
with
\begin{equation}
	\mbox{\bf A} (\mbox{\bf R}_m) 
		= \sum_n q_n \mbox{\bf a} (\mbox{\bf R}_m (\tau) 
		- \mbox{\bf r}_n (\tau)) \; ,
\end{equation}
and
\begin{equation}
	\mbox{\bf a} (\mbox{\bf r}) = 
		- \frac{1}{8a^2} \frac{C_{\rm x}}{C_1} \left[1 +
		2\ln\left(\frac{a}{r}\right)\right] 
	[\hat z \times \mbox{\bf r}] \; .
\end{equation}
The charges are the sources of a vector potential felt by the 
moving vortices. It depends of the signs of the corresponding 
vortices and charges. 
We can  rewrite this term, after a partial integration in
Eq.(\ref{Sqv2}), in such a form that the vortices create a 
gauge potential for the charges. Hence this charge-vortex 
interaction term fully preserves the duality between charges
and vortices.

In summary the resulting action 
\begin{equation} \label{fin}
	S\{\mbox{\bf r},\mbox{\bf R}\} 
	= S_{\rm int}^{\rm (q)} + S_{\rm int}^{\rm (v)} +
	S_{\rm kin}^{\rm (q)} + S_{\rm kin}^{\rm (v)} + S_{\rm qv}  
\end{equation}
is essentially that of two coupled 2D Coulomb gases.
It is symmetric with respect to charges and vortices. Both
can be considered as particles with masses 
\begin{equation} \label{mass1}
	M_{\rm q} =  \frac{1}{a^2 E_{\rm J1}} \ \ \mbox{and} \; \;
	M_{\rm v} =  \frac{\pi^2}{4a^2 E_{\rm C2}} \ \,
\end{equation}
respectively. Charges interact via the effective capacitance, vortices
via the usual logarithmic interaction with strength $E_{\rm J2}$. In
addition, there is a specific feature of two-layered system:  
the vortices produce a vector potential $\mbox{\bf a}$ for the
charges, whereas charges create a gauge potential
for vortices. The magnetic field, associated with this vector
potential is  
\begin{equation} \label{field}
	B = \pm \frac{1}{2ea^2} \frac{C_{\rm x}}{C_1} 
	\ln \frac{a}{r},\ \ \ a \ll r \ll \xi_1.
\end{equation}
 The interaction described by this vector/gauge potential
is always controlled by the small parameter $C_{\rm x}/C_1$. In the
regime considered it is weak.

\section{Magnetic field -- offset charge duality}

In this last section we return to the case of a single Josephson 
junction array. We assume that the array is in a perpendicular
position-dependent
magnetic field ${\bf B}_i$. Furthermore, we allow for ``offset''
charges on the grains $Q_{{\rm x}i}$, which 
in principle can be  controlled by applied gate voltages between the 
ground and the islands.
(In practice they are caused e.g. by random impurity charges in 
the substrate.) Both these factors can be accounted in the formalism
described above. 
The action describing this system is (see e.g. \cite {AvO})\\
\\
$
	S\{\phi,q\} = 
$
$$
\int_{0}^{\beta} d\tau \left\{ \vphantom{\frac{}{{}_{}}} \right. 
	2e^2 \sum_{ij}
	[q_i(\tau) + q_{{\rm x}i}] C^{-1}_{ij} [q_j(\tau) + q_{xj}] 
$$
\vspace{-0.4cm}
\begin{eqnarray} \label{S1}
 + i \sum_i
	q_i(\tau) \dot\phi_i (\tau) 
 - E_{\rm J}\sum_{\langle ij \rangle} \cos
	[\phi_{i} - \phi_{j} - A_{ij}] 
	\left.  \vphantom{\frac{}{{}_{}}} \right\}. 
\end{eqnarray}
Here $q_{{\rm x}i} = Q_{{\rm x}i}/2e$, and the magnetic field is taken into
account by the vector-potential $A_{ij} = (2e/c)\int_i^j
{\bf A} \cdot {\bf dl}$. After the same transformations as those
performed in Sec. 2, we arrive to the coupled-Coulomb gas action for
charges and vorticities, similar to Eq. (\ref{Sqv11})\\
\\
$
	S\left\{ q, v \right\} =  
$
$$
	\int_0^{\beta} d\tau \sum_{ij} \left\{ 2e^2 [q_i(\tau) + q_{{\rm x}i}]
	C^{-1}_{ij} [q_j(\tau) + q_{{\rm x}j}] \frac{}{{}_{}} \right.
$$
$$
	+ \pi E_{\rm J} [v_i(\tau) + f_i] G_{ij}
	[v_j(\tau) + f_j] 
$$
\begin{eqnarray} \label{S2}
\left. + i[q_i(\tau) + q_{{\rm x}i}] \Theta_{ij} 
	\dot{v}_j(\tau) +
	\frac{1}{4\pi E_{\rm J}} \dot{q}_i(\tau) G_{ij} 
					\dot{q}_j(\tau) \right\}.
\end{eqnarray}
The ``frustration''  
$$
	f_i = 
	(2\pi)^{-1} \epsilon^{(\mu \nu)} \nabla_{\nu} A_{i,i+\mu}
$$ 
describes the magnetic flux through the plaquette $i$, measured in
units of the flux quantum $\Phi_0 = \pi c/2e$. Note that due to the
fact that the charges and vorticities are integer-valued, only the
fractional parts of both offset charges $q_{{\rm x}i}$ and frustrations $f_i$
matters. 

The presence of offset charges $Q_{{\rm x}i}=2eq_{{\rm x}i}$, 
breaking the particle-hole
symmetry, has profound consequence on the vortex motion.
Offset charges at positions $\vec{r}_i$ are responsible for a gauge potential
\begin{equation}
	\vec{A}(\vec{r})
	= \sum_{i}q_{{\rm x}i}\vec{\nabla}\Theta(\vec{r}-\vec{r}_{i}) \; ,
\label{Aq}
\end{equation}
which acts on the vortices in the same way as an ordinary vector
potential acts on charges. 
Associated with this `vector potential' $\vec{A}(\vec{r})$ 
is a `magnetic field' and a `Lorentz force' acting on moving
vortices. We call this force a {\sl Magnus force}. 
A vortex with vorticity $v_n$ and velocity $\dot{\vec{R}}$
feels the Magnus force~\cite{AvO}
\begin{equation}
	\vec{F}_{\rm M} =  v_n q_{\rm x} \hat{z} \times \dot{\vec{R}} \; .
\label{Magnus}
\end{equation}
Here we assumed for simplicity a homogeneous gate charge.
As a result of the combined effect of the Magnus force (\ref{Magnus}) and
the Lorentz force, which is the force
on the vortex due to an external feeding current, the vortices
will move at a certain angle with respect to the current.
This angle is called Hall angle. Its measurement yields information on
the different forces in the system.

Real samples are usually characterized by random offset charges. As a
result the Magnus force averages to approximately zero. This effect
is probably responsible for the small size of the observed Hall angle in 
Josephson junction arrays. The forces on vortices
 in Josephson junction arrays
have been discussed recently in the literature, in part
in a controversial way (see Ref. \cite{Gaitan,Zhu,MV,Sonin,OSB}).
Here we want to stress that only the offset charges, which are
responsible for a local deviations from charge neutrality in the array,
lead to the Magnus force.

\section{Conclusions}

We have considered two-dimensional Josephson junction
arrays. In the classical limit, $E_{\rm J} \gg E_{\rm C}$, the
vortices are 
the relevant excitations. In the opposite quantum limit $E_{\rm C} \gg
E_{\rm J}$ the charges of the islands are well-defined degrees of
freedom. Both can be described simultaneously by a coupled-Coulomb-gas
action, which is  dual under the interchange of  charges and  vortices.
In each of the two limiting cases ($E_{\rm J} \ll/\gg E_{\rm C}$)
the system reduces to a one-component Coulomb gas, where 
either the charges or the vortices  can
be considered as (logarithmically) interacting massive quantum
particles. Furthermore, the external magnetic field plays the same
role for charges, as the external (``offset'') charges for vortices. 

The situation is even more interesting in a system of two parallel,
 capacitively coupled Josephson junction arrays. It is described by
a coupled-Coulomb-gas action for charges and vortices in both
arrays. If one array is in the semi-classical (vortex)
regime, while the other is in the quantum (charge) regime
we can  integrate out the strongly fluctuating  variables. In this
 case we 
arrive at an effective action, describing {\em dual} charges in one 
array and vortices in  the other, which both are now dynamic degrees of
freedom, in contrast to the one-layer problem. Furthermore, vortices
and charges interact via gauge field of strength  controlled
by the interlayer capacitance. This 
peculiar interaction  between charges and
vortices resembles the composite fermion scenario of the
fractional quantum Hall effect, which may become a subject of future
research.   

\section*{Acknowledgments}

The work was supported by  the Deutsche Forschungsgemeinschaft within the
research program of SFB 195 and by the Swiss National Science Foundation
(Y.~M.~B.). 

\section*{Appendix}

In this Appendix we provide the steps leading from
Eqs. (\ref{part11}), (\ref{Sqphi5}) to Eqs. (\ref{part12}),
(\ref{Sqv11}). 
Starting from the partition function (\ref{part11}), we first introduce
the vortex degrees of freedom. This can be done by means of the Villain
transformation~\cite{Villain} (see also \cite{JKKN}); the
time-dependent quantum problem requires some additional steps
\cite{Fazio,FGS}. We introduce the lattice with spacing $\epsilon$ in
time direction; this spacing is of order of inverse Josephson
frequency: $\epsilon \sim (8E_{\rm J}E_{\rm C})^{1/2}$. In the Villain
approximation one replaces 
$$
	\exp \Big\{ -\epsilon E_{\rm J} 
	\sum_{\langle i j \rangle, \tau} 
	[1 - \cos(\phi_{i,\tau} - \phi_{j,\tau})] \Big\} 
$$
$$
\rightarrow \sum_{\{
	{\bf m}_{i\tau}\}} \exp \Big\{ -\frac{\epsilon E_{\rm J} F(\epsilon
	E_{\rm J})}{2} \sum_{i,\tau} \vert \nabla \phi_{i\tau} - 2\pi
	{\bf m}_{i\tau} \vert^2 \Big\}. \eqno(A1)
$$  
Here we have introduced a two-dimensional vector field
${\bf m_{i\tau}}$, defined on dual lattice (alternatively, it can be
considered as a scalar field defined on bonds). The function
$$
	F(x) = \frac{1}{2x \ln({J_0(x)/J_1(x)})} \to \frac{1}{2x \ln(4/x)}, \ \
\ x \ll 1,$$ is introduced to ``correct'' the Villain transformation
for small $E_{\rm J}$ (see e.g. Ref. \cite{Otterlo}). 
As we see, its entire effect is to
renormalize (increase) the Josephson coupling $E_{\rm J} \to E_{\rm J}
F(\epsilon E_{\rm J})$, but it does not affect the physics. In
the following we will use only the renormalized constant. 

The rhs.\ of Eq. (A1) can be rewritten as
$$
	\sum_{\{ {\bf J}_{i\tau}\}} \exp 
	\Big\{ -\frac{1}{2\epsilon E_{\rm J}}
	\sum_{i,\tau} \vert {\bf J}_{i\tau} \vert^2 - i {\bf J_{i\tau}}
	\nabla \phi_{i\tau} \Big\} \eqno(A2)
$$
Now the Gaussian integration over the phases can be easily performed,
yielding 
$$
	Z = \sum_{q_{i\tau}} \sum_{{\bf J}_{i\tau}} \exp \Big\{ -2e^2
	\epsilon \sum_{i,j,\tau} q_{i\tau}  C^{-1}_{ij}
	q_{j\tau} 
$$
\vspace{-0.3cm}
$$
\;\;\;\;\;\;\;\;\;\;\; - \frac{1}{2\epsilon E_{\rm J}} \sum_{i,\tau} \vert
	{\bf J}_{i\tau} \vert^2 \Big\}, \eqno(A3)
$$
and the summation is constrained by the continuity equation, 
$$
	\nabla {\bf J}_{i\tau} - \dot{q}_{i\tau} = 0 \;.
$$
The time derivative stands for a discrete derivative
$\dot{f} (\tau) = \epsilon_{\mu}^{-1} [f(\tau + \epsilon_{\mu}) -
f(\tau)]$.
The constraint is satisfied by the parameterization \cite{Elitzur}
$$
	J^{(\mu)}_{i\tau} = 
	n^{(\mu)} ({\bf n}\nabla)^{-1} \dot{q}_{i\tau} +
	\epsilon^{(\mu\nu)} \nabla_{\nu} A_{i\tau}.
$$ 
Here the operator $({\bf n}\nabla)^{-1}$ is the line integral on the
lattice (in Fourier space it has the form $i(k_x + k_y)^{-1}$),
$\epsilon^{(\mu \nu)}$ is the antisymmetric tensor, while
$A_{i\tau}$ is an unconstrained integer-valued scalar field. 

With the use
of the Poisson resummation (which requires introducing  a new
integer scalar field $v_{i\tau}$) the partition function can be
rewritten as
$$
	Z = \sum_{q_{i\tau},v_{i\tau}} \exp - S \{q,v\} \;.
$$
The effective action for the integer charges $q_{i}$ and vorticities
$v_{i}$ is\\
\\
$
S\{q,v\} = \Bigg\{ 2e^2 \epsilon \sum_{ij\tau} q_{i\tau}
	C^{-1}_{ij} q_{j\tau} $
\vspace{-0.3cm}
$$
\;\;\;\;\;\;\;\;\;\;\;
 - \frac{1}{2\epsilon E_{\rm J}} 
	\sum_{i\tau} \Big[ n^{(\mu)}
	({\bf n} \nabla)^{-1} \dot{q}_{i\tau} \Big]^2   
$$
\vspace{-0.3cm}
$$       -\; \frac{\epsilon
	E_{\rm J}}{4\pi} \sum_{ij\tau} \Big[ 2\pi v_{i\tau} -
	\frac{i}{\epsilon E_{\rm J}} 
	\epsilon^{(\mu \nu)} \nabla_{\nu} n^{(\mu)}
	({\bf n} \nabla)^{-1} \dot{q}_{i\tau} \Big] G_{ij}
$$
\vspace{-0.2cm}
$$
\;\;\;
	\times \Big[ 2\pi
	v_{i\tau} - \frac{i}{\epsilon E_{\rm J}} 
	\epsilon^{(\mu \nu)} \nabla_{\nu}
	n^{(\mu)} ({\bf n} \nabla)^{-1} \dot{q}_{j\tau} \Big]
	\Bigg\}.	\eqno(A4)
$$
The kernel $G_{ij}$ is the lattice Green's function,
i.e. the Fourier transform of $k^{-2}$. Finally, after some algebra
\cite{FGS} we arrive to the effective action (\ref{Sqv11}), rewritten in
the continuous notations.


\begin{thebibliography}{99}
\bibitem{Foot1} It is easy to show that the vortices with $\vert v_n
\vert >1$ are unstable and decay to separate vortices with $v_n = \pm
1$.
\end{thebibliography}
\end{document}